# Enhancing Price Prediction in Cryptocurrency Using Transformer Neural Network and Technical Indicators

Mohammad Ali Labbaf Khaniki*, Mohammad Manthouri

Faculty of Electrical Engineering, K.N. Toosi University of Technology, Tehran, Iran

*<mohammadlabbaf@email.kntu.ac.ir>

**Abstract:** This study presents an innovative approach for predicting cryptocurrency time series, specifically focusing on Bitcoin, Ethereum, and Litecoin. The methodology integrates the use of technical indicators, a Performer neural network, and BiLSTM (Bidirectional Long Short-Term Memory) to capture temporal dynamics and extract significant features from raw cryptocurrency data. The application of technical indicators, such facilitates the extraction of intricate patterns, momentum, volatility, and trends. The Performer neural network, employing Fast Attention Via positive Orthogonal Random features (FAVOR+), has demonstrated superior computational efficiency and scalability compared to the traditional Multi-head attention mechanism in Transformer models. Additionally, the integration of BiLSTM in the feedforward network enhances the model's capacity to capture temporal dynamics in the data, processing it in both forward and backward directions. This is particularly advantageous for time series data where past and future data points can influence the current state. The proposed method has been applied to the hourly and daily timeframes of the major cryptocurrencies and its performance has been benchmarked against other methods documented in the literature. The results underscore the potential of the proposed method to outperform existing models, marking a significant progression in the field of cryptocurrency price prediction.

**Keywords:** Cryptocurrency, Deep Learning, Time Series prediction, Transformer, Performer, Attention Mechanism,

1) **Introduction**

In the rapidly evolving landscape of technology, the mode of transactions has undergone a significant paradigm shift. Traditional physical payments, such as cash and cheques, are increasingly being replaced by digital transactions. This transformation has been largely driven by the advent and proliferation of cryptocurrencies, which have emerged as a new asset class and medium of exchange (Aghashahi and Bamdad, 2023). Cryptocurrencies, unlike conventional fiat currencies, employ cryptographic ciphers to facilitate financial transactions. Over the past decade, digital finance has witnessed exponential growth, with cryptocurrencies leading this innovative stride forward. One of the most critical aspects of using any currency, whether as a medium of transaction or as an asset, is the ability to predict its expected value. The value and stability of any currency largely depend on the controlling authority. In the case of fiat currencies, this is typically the government of the respective country. However, cryptocurrencies operate in a decentralized manner, free from governmental control. This unique characteristic presents both opportunities and challenges. On one hand, it offers potential for high returns and diversification. This underscores the need for robust and accurate price prediction models in the realm of cryptocurrencies (Pichaiyuth *et al.*, 2023).

Predicting the price of cryptocurrencies, such as Bitcoin, presents a unique set of challenges. The primary difficulty lies in the inherent volatility of the cryptocurrency market. Unlike traditional financial markets, the cryptocurrency market is open 24/7, leading to more frequent price changes. Furthermore, the price of cryptocurrencies can be influenced by a variety of factors, including technological advancements, regulatory news, market sentiment, and macroeconomic trends (Zhao, Crane and Bezbradica, 2022). Additionally, the lack of a centralized authority adds another layer of complexity, as the value is not tied to a physical asset or controlled by a single entity. These factors contribute to the unpredictability of cryptocurrency prices, making accurate prediction a complex task. Despite these challenges, the enormous potential value of cryptocurrencies has attracted significant attention from both investors and researchers, leading to the development of sophisticated prediction models (Awoke *et al.*, 2021).

Technical indicators play a crucial role in cryptocurrency trading and analysis. They provide traders with a statistical approach to assess market conditions and forecast price trends. Indicators such as the Relative Strength Index (RSI) and moving averages are commonly used to identify potential buy or sell signals. For instance, RSI can help determine whether a cryptocurrency is overbought or oversold, while moving averages can signal the start of a bullish or bearish trend. Moreover, these indicators can be used in conjunction with machine learning models to enhance prediction accuracy (Pichaiyuth *et al.*, 2023). By extracting meaningful features from raw price data, they enable these models to capture complex patterns and trends, thereby improving the effectiveness of cryptocurrency price prediction. These technical indicators can serve as valuable inputs to predictive models. By capturing key statistical properties of the market, they can help the models identify potential future movements in cryptocurrency prices (Goutte *et al.*, 2023).

Transformers, a neural network architecture, have garnered considerable interest in both Natural Language Processing (NLP) and time series analysis. Their capacity to manage long-range dependencies and parallel processing has propelled their popularity in these domains (Rahali and Akhloufi, 2023). Utilizing Multi-head self-attention or scaled dot-product attention, Transformers can assess the significance of inputs within a sequence, enabling the capture of intricate data patterns. In NLP, Transformers have been pivotal in achieving cutting-edge performance across tasks like translation, summarization, and sentiment analysis (Patwardhan, Marrone and Sansone, 2023). In time series prediction, Transformers exhibit promise, leveraging their capability to grasp temporal relationships. Applications range from weather and cryptocurrency market forecasting to fault and anomaly detection. It can identify patterns and trends over time, making it possible to predict future data points with a high degree of accuracy (Haryono, Sarno and Sungkono, 2023).

The attention mechanism plays a pivotal role in the Transformer architecture, significantly enhancing its performance in time series prediction (Vaswani *et al.*, 2017). The attention mechanism allows the model to focus on different parts of the input sequence when producing an output, effectively capturing the dependencies between words or events that are far apart. In time series prediction, the attention mechanism allows the Transformer to weigh the importance of past events when predicting future ones. This is particularly useful in scenarios where recent events may not be the most relevant for making a prediction. Moreover, the attention mechanism in Transformers is computationally efficient as it allows for parallel computation across the sequence, unlike RNNs (Recurrent Neural Network) which require sequential computation (Samii *et al.*, 2023). This makes Transformers faster and more scalable for large datasets (Labbaf Khaniki, Mirzaeibonehkhater and Manthouri, 2023).

This research introduces a pioneering methodology for time series prediction of Bitcoin, Ethereum, and Litecoin. The approach initially utilizes technical indicators to extract statistical features from the data. Following this, a Performer neural network is applied, which uses Fast Attention Via positive Orthogonal Random features (FAVOR+) instead of Multi-head attention mechanism used in Transformer. The architecture further includes BiLSTM, enhancing its ability to capture temporal dynamics. The key advancements of this research will be elaborated in the following sections. The key advancements of this method are outlined below:

1. **Feature Extraction using Technical Indicators**: In the realm of financial analysis, technical indicators such as RSI and moving averages are often used to predict future price movements based on historical data. By using these indicators, the research extracts meaningful features from the raw cryptocurrency data, which can capture complex patterns and trends that might be missed by the naked eye.

2. **Performer**: In this research takes this a step further by incorporating a Performer neural network. The FAVOR+ mechanism is more computationally efficient and scalable than the Multi-head attention mechanism. This is because it approximates the attention mechanism in a way that requires less computational resources, making it possible to process larger sequences of data. This is particularly beneficial in tasks like time series prediction, where the model needs to process long sequences of historical data to make accurate predictions.

3. **BiLSTM**: The use of BiLSTM (Bidirectional Long Short-Term Memory) in the feedforward network allows your model to capture temporal dynamics in the data, as it can process the data in both forward and backward directions. This is particularly useful for time series data where past and future data points can influence the current state. The fully connected layers, on the other hand, enable the model to learn complex non-linear relationships between the features. This combination of BiLSTM and fully connected layers makes your model capable of handling the complexity and volatility often seen in cryptocurrency price movements.

The novel components of this research collectively form an advanced method for predicting time series for cryptocurrencies, potentially surpassing the predictive performance of existing models. To evaluate our approach, we apply it to the hourly and daily timeframes of major cryptocurrencies such as Bitcoin, Ethereum, and Litecoin. We then compare its performance with other methods documented in the literature, specifically those presented in (Awoke *et al.*, 2021), (Jay *et al.*, 2020), and (Girsang and Stanley, 2023).

The organization of this paper is as follows: Section II offers a review of the relevant literature. Section III delves into the proposed methodology, discussing the use of financial indicators for feature extraction and the proposed deep learning method that combines the BiLSTM and Performer. Section IV showcases the simulation results, providing details on the training and evaluation process of the proposed approach. Finally, Section V wraps up the paper, summarizing the key findings and contributions of our study.

**2) Related Works**

Over the past few decades, machine learning and deep learning have brought about substantial changes in the field of financial forecasting, including the prediction of cryptocurrency prices (Zhang *et al.*, 2021) and (Mohammadabadi *et al.*, 2023). Machine learning models utilize historical price data to anticipate future trends, effectively capturing intricate, non-linear patterns in the data. Deep learning, an advanced branch of machine learning, employs multi-layered neural networks (McCarthy *et al.*, 2020) and (McCarthy *et al.*, 2021). These models are particularly adept at dealing with time series data, such as cryptocurrency prices, by modeling intricate patterns and interdependencies in the data. In this section, we conduct a review of the existing literature on the prediction of cryptocurrency prices using machine learning and deep learning techniques.

In the initial stages, the most recognized model was the moving average autoregressive model, ARIMA (Abu Bakar and Rosbi, 2017). Subsequently, (Kim, Jun and Lee, 2021) utilized the GARCH model (autoregressive conditional heteroskedasticity model) for forecasting the cryptocurrency market data. While these techniques can be effectively used for short-term prediction, they are not suitable for nonlinear problems and exhibit poor long-term prediction performance (Fang *et al.*, 2023) and (Safari, Khalfalla and Imani, no date). To address this issue, machine learning was introduced to analyze time series and has been successfully applied to cryptocurrency price forecasting. The ability of machine learning to process complex and large volumes of data has resolved many limitations of traditional methods and. Machine learning methods include Support Vector Machine (SVM), decision tree, naive Bayes, random forest (Rathan, Sai and Manikanta, 2019). (Orte *et al.*, 2023) combined decision trees and SVM models to predict future price trends. (Cortez, Rodríguez-García and Mongrut, 2020) developed a feature-weighted SVM and K-nearest neighbor algorithm to predict the cryptocurrency market index. Experimental results have demonstrated that the model has good short-term, medium-term, and long-term prediction capabilities (Rabiee and Safari, 2023) and (Safari, Khalfalla and Imani, 2022).

In recent years, deep learning techniques that solely depend on datasets have been utilized to predict cryptocurrency prices, eliminating the need for expert knowledge (Safari, Khalfalla and Imani, 2022). Consequently, their use in cryptocurrency prediction has increasingly become a focal point of scholarly research. Deep learning methods encompass Gated Recurrent Unit (GRU), RNN, Convolutional Neural Network (CNN), LSTM, and BiLSTM (Wegayehu and Muluneh, 2022), (Omran *et al.*, 2021), (Seabe, Moutsinga and Pindza, 2023). In, CNN was sequentially employed for cryptocurrency price prediction (Ramadhani *et al.*, 2018; Alonso-Monsalve *et al.*, 2020). In (Seabe, Moutsinga and Pindza, 2023), a Conv1D-LSTM model was proposed, which merges one-dimensional CNN and LSTM. This combination leverages the strengths of both networks. (Ramakrishnan *et al.*, 2022) carried out predictive research on global cryptocurrency indexes using BiLSTM, demonstrating that BiLSTM possesses excellent predictive accuracy and robust generalization capability.

The attention mechanism is a key component of the Transformer architecture, providing it with the ability to focus on different parts of the input sequence when generating predictions. This feature is particularly beneficial in time series prediction as it allows the model to weigh the importance of past and recent events differently, thereby significantly improving the accuracy of the predictions. The authors in (Totaro, Hussain and Scardapane, 2020) demonstrate the application of this technique in time series forecasting with the dual-stage attention-based recurrent neural network. This model is applied to the hourly data of Dogecoin price for its prediction over time. The paper (Zhang *et al.*, 2021) introduces a novel approach for predicting cryptocurrency prices. The authors propose a model that combines CNNs with weighted and attentive memory channels. This unique combination allows the model to effectively extract features from the data and capture complex patterns.

Researchers have proposed adaptations like the temporal Fusion Transformer and time-series Transformer to optimize its architecture for time-series data. (Lim *et al.*, 2021) presents another adaptation of Transformer, called the temporal fusion Transformer, that combines high-dimensional and diverse inputs from multiple sources to produce accurate and interpretable forecasts for various time horizons. The paper (Sridhar and Sanagavarapu, 2021) presents a novel approach to predicting Dogecoin prices using a Multi-head self-attention Transformer. (Li *et al.*, 2019) proposes a novel variant of Transformer, called the time-series Transformer, that improves the performance and efficiency of Transformer on time series forecasting tasks. The paper (Son *et al.*, 2022) presents a novel approach to predict cryptocurrency prices by analyzing social media trends. The paper (You *et al.*, 2022) introduces a novel spatiotemporal Transformer designed to predict high-dimensional short-term time-series data. This approach model leverages a spatiotemporal information transformation equation and a continuous attention mechanism to enhance prediction accuracy. The paper (Tanwar and Kumar, 2022) explores the application of advanced deep learning techniques to forecast the prices of cryptocurrencies. It specifically examines the use of Transformers in conjunction with LSTM networks to analyze financial time series data for cryptocurrencies like Ethereum and Bitcoin. The paper (Yunsi, Lahcen and Azzouz Mohamed, 2023) explores the application of Transformer neural networks to predict cryptocurrency prices. The paper (Du, Côté and Liu, 2023) proposes a novel method based on the self-attention mechanism containing of two diagonally-masked self-attention blocks that learn missing values from a weighted combination of temporal and feature dependencies.

Upon examining the literature review, it's evident that the Transformer neural network structure has seen a surge in usage recently, compared to earlier deep learning models. This architecture was designed to solve sequence-to-sequence tasks while handling long-range dependencies with ease. The key advantage of the Transformer over previous architectures like RNNs is its ability to handle long-range dependency. In simple terms, it has a stronger memory when it comes to remembering old connections.

**3) Methodology**

This section provides an in-depth examination of the Proposed Performer Neural Network and its components, which include Technical Analysis, the Multi-head Attention Mechanism in Transformer, Performer, and BiLSTM. It introduces technical indicators as statistical features that aid in the analysis of financial data. The Multi-head Attention Mechanism, a key element of the transformer architecture, is explained in terms of its functionality and underlying formula. The Performer, an efficient variant of the transformer, is described along with its relevance to the model. Lastly, the section covers LSTM and BiLSTM, which are types of RNNs used in the feedforward network of the proposed method, highlighting their role in capturing temporal dependencies in the data.

**3.1) Technical Analysis**

Technical indicators provide a quantified measure of market conditions and trends, which can serve as valuable input features for a deep learning model. Technical indicators, when used as features in deep learning models for Bitcoin price forecasting, can significantly enhance the model's predictive capabilities. These indicators encapsulate key market trends and behaviors, providing a rich, quantified dataset from which the model can learn. By integrating indicators, deep learning models can capture complex, non-linear relationships in the data that might be missed by traditional analysis. For instance, a deep learning model can use these indicators to identify patterns that precede market movements, allowing for more accurate predictions of price changes. The model can learn from historical data how certain indicator values correlate with upward or downward trends in Bitcoin prices. This learning enables the model to anticipate similar movements when these indicator patterns reoccur. Moreover, technical indicators can serve as a normalization tool, helping to scale and transform the input data into a format that is more digestible for the model, thus improving the training process. By providing a standardized input, the model can focus on the underlying patterns rather than getting confused by noise or scale differences in raw price data (Wang *et al.*, 2023).

In this study, several technical indicators are utilized, including the simple moving average, exponential moving average, Bollinger Bands, RSI, and Commodity Channel Index (CCI). The corresponding mathematical formulas for these indicators are also provided.

1. **Simple Moving Average (SMA)**: The SMA is calculated by adding the price of an instrument over a number of time periods and then dividing the sum by the number of time periods. The formula is:

$$SMA = \frac{1}{n} \sum_{i=1}^{n} P_i \qquad (1)$$

where $P_i$ is the price at period $i$ and $n$ is the number of periods.

2. **Exponential Moving Average (EMA)**: The EMA is a type of weighted moving average that gives more importance to the latest data. The formula is:

$$EMA_i = (P_i - EMA_{i-1}) \times k + EMA_{i-1} \qquad (2)$$

where $P_i$ is the current price, $EMA_{i-1}$ is the EMA value for the previous price, and $k$ is $\frac{2}{(number\ of\ periods)}$.

3. **Bollinger Bands (BB)**: The BB consist of a middle band with two outer bands. The middle band is a simple moving average, and the outer bands are standard deviations away from the middle band. The formulas are:

$$Middle\ Band = SMA(n) \qquad (4)$$

$$Upper\ Band = SMA(n) + k \times std(n) \qquad (5)$$

$$Lower\ Band = SMA(n) - k \times std(n) \qquad (6)$$

where n is the number of periods, and $k$ is a constant (usually 2).

4. **Relative Strength Index (RSI)**: The RSI compares the magnitude of recent gains to recent losses in an attempt to determine overbought and oversold conditions of an instrument. The formula is:

$$RSI = 100 - \frac{100}{1 + RS} \tag{7}$$

where $RS$ is the average gain over $n$ periods divided by the average loss over $k$ periods.

5. **Commodity Channel Index (CCI)**: The CCI measures the difference between a security's price change and its average price change. High positive readings indicate that prices are well above their average, which is a show of strength. The formula is:

$$CCI = \frac{(TP - MA)}{(0.015 \times D)} \tag{8}$$

where $TP$ is the typical price $TP = \frac{(High + Low + Close)}{3}$, MA is the moving average, and $D$ is the mean deviation.

### 3.2) Multi-head Attention Mechanism

The Transformer model's core is the Multi-head attention mechanism, which is pivotal for handling sequential data, especially in natural language processing. It enables the model to concurrently pay attention to various segments of the input, detecting intricate patterns and interdependencies. Within this framework, Multi-head attention plays a crucial role in discerning complex connections over different intervals in a sequence, thus improving the model's proficiency in identifying temporal links and significant elements.

The components of the attention mechanism are:

- Queries (Q): The targeted information to be retrieved from the input.

- Keys (K): They provide the necessary context for the sequence's elements.

- Values (V): They contain the actual data related to each sequence element.

During the attention phase, queries are matched against keys to ascertain the amount of data to be drawn from each value. The Attention score is computed as follows.

$$Attention(Q, K, V) = Softmax\left(\frac{Q \cdot K^T}{\sqrt{d_k}}\right) \cdot V \tag{9}$$

In the given formula, $d_k$ denotes the key vectors' size. The square root of $d_k$, $\sqrt{d_k}$, is used to scale the computations. During self-attention, where queries, keys, and values originate from identical sequences, this scaling helps the model to optimally allocate information among the sequence's various elements.

Multi-head attention is a key component of the Transformer architecture, which allows the model to focus on different parts of the input sequence simultaneously. Each attention head processes the input independently, allowing the model to capture different types of information from the same input sequence. The following formulas are for the Multi-head Attention mechanism in Transformer models. The Multi-head Attention function is defined as follows:

$$Multi - Head(Q, K, V) = \text{Concat}(head_1, \dots, head_h)W_o, \tag{10}$$

In this equation, each $head_i$ represents an individual attention head, and $W_o$ is the output weight matrix. The Concat function concatenates the output of all attention heads, and this result is then multiplied by the output weight matrix $W_o$.

Each individual attention head ($head_i$) is computed as follows:

$$head_i = Attention(QW_{Qi}, KW_{Ki}, VW_{Vi}) = Softmax\left(\frac{(QW_{Qi}) \cdot (KW_{Ki})^T}{\sqrt{d_k}}\right) \times VW_{Vi} \quad (11)$$

In this equation, $QW_{Qi}$, $KW_{Ki}$, and $VW_{Vi}$ are the query, key, and value matrices for the i-th attention head, respectively. These matrices are obtained by multiplying the input query, key, and value matrices ($Q$, $K$, and $V$) with their corresponding weight matrices ($QW_{Qi}$, $KW_{Ki}$, and $VW_{Vi}$). The Softmax function is applied to the dot product of the query and key matrices, scaled by the square root of the dimensionality of the keys $\sqrt{d_k}$. This result is then element-wise multiplied by the value matrix $VW_{Vi}$. These formulas allow the model to process input with multiple attention heads, each capturing different types of information from the same input sequence. The block diagram of the Multi-head attention mechanism is shown in Fig. 1.

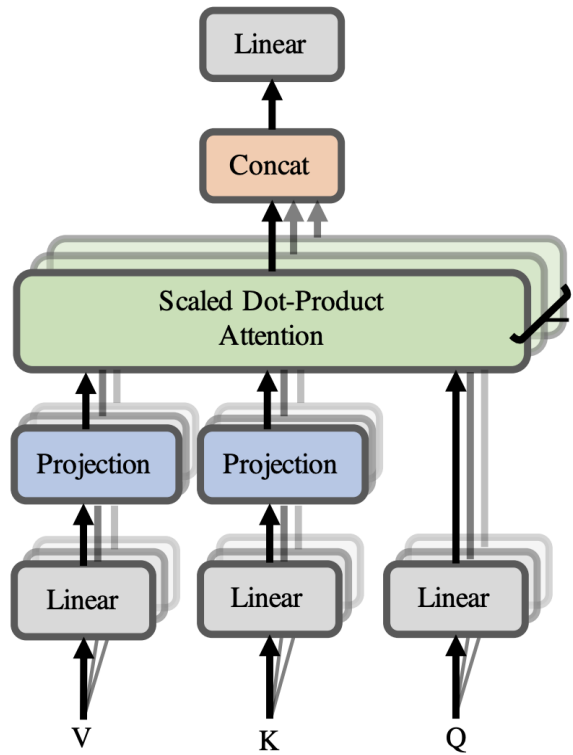

Fig. 1. The block diagram of Multi-head attention mechanism

### 3.3) Performer

The paper "Rethinking Attention with Performers" introduces Performers, a novel Transformer architecture that can estimate regular full-rank-attention Transformers with provable accuracy, but with only linear space and time complexity (Wang *et al.*, 2023). The Performers use a new approach called Fast Attention Via Positive Orthogonal Random features (FAVOR+), which can also be used to efficiently model kernelizable attention mechanisms beyond Softmax. This allows for accurate comparison of Softmax with other kernels on large-scale tasks, beyond the reach of regular Transformers. The Performers were tested on a variety of tasks, from pixel prediction to text models to protein sequence modeling, and demonstrated competitive results with other efficient sparse and dense attention methods (Choromanski *et al.*, 2020).

Consider a time series with length L. The standard dot-product attention, is a mapping that takes matrices $Q$, $K$, and $V \in \mathbb{R}^{L \times d}$ as input, where d is the dimension of the hidden state (the dimension of the latent representation). These matrices $Q$, $K$, and $V$ are intermediate representations of the time series, and their rows can be seen as queries, keys, and values of the continuous dictionary data structure, respectively. The bidirectional dot-product attention takes the following form, where $A \in \mathbb{R}^{L \times L}$ is the so-called attention matrix:

$$Attention_{\hookleftarrow}(Q, K, V) = D^{-1}AV, \tag{12}$$

$$A = exp\left(\frac{Q \cdot K^T}{\sqrt{d_k}}\right), \tag{13}$$

$$D = diag\ (A1_L), \tag{14}$$

here, the $exp(\cdot)$ function is applied elementwise, $1_L$ is an all-ones vector of length $L$, and $diag(\cdot)$ is a diagonal matrix with the input vector as the diagonal. The time and space complexity of computing (1) are $O(L^2 d)$ and $O(L^2 + Ld)$ respectively, because $A$ has to be stored explicitly. Therefore, in principle, dot-product attention of type (1) is incompatible with end-to-end processing of long time series. Bidirectional attention is used in encoder self-attention and encoder-decoder attention in Seq2Seq architectures. Another significant type of attention is unidirectional dot-product attention, which takes the form:

$$Attention_\hookrightarrow(Q, K, V) = \tilde{D}^{-1}\tilde{A}V, \tag{15}$$

$$\tilde{A} = tril(A), \tag{16}$$

$$\tilde{D} = diag\left(\tilde{A}1_L\right), \tag{17}$$

where $tril(\cdot)$ returns the lower-triangular part of the argument matrix, including the diagonal. The unidirectional attention is used for autoregressive generative modeling, e.g., as self-attention in generative Transformers as well as the decoder part of Seq2Seq Transformers. FAVOR+ operates for attention blocks using matrices $A \in \mathbb{R}^{L \times L}$ of the form $A(i,j) = K(q_i^T, k_i^T)$, with $q_i/k_j$ representing the $i^{th}/j^{th}$ query/key row-vector in $Q/K$ and kernel $K: \mathbb{R}^d \times \mathbb{R}^d \to \mathbb{R}_+^d$ defined for the (usually randomized) mapping: $\varphi: \mathbb{R}^d \to \mathbb{R}_+^d$ (for some $r > 0$) as:

$$K(x, y) = \mathbb{E}\left[\varphi(x)^T \varphi(y)\right]. \tag{18}$$

We refer to $\varphi(u)$ as a random feature map for $u \in \mathbb{R}^d$. For $\acute{Q}, \acute{K} \in \mathbb{R}^L$ with rows given as $\varphi(q_i^T)^T$ and $\varphi(k_i^T)^T$ respectively, Equation (19) leads directly to the efficient attention mechanism of the form:

$$\widehat{Attention}_\hookrightarrow(Q, K, V) = \widehat{D}^{-1}\left(\acute{Q} \cdot (\acute{K})^T \cdot V\right), \tag{19}$$

$$\widehat{D}^{-1} = diag\left(\acute{Q} \cdot (\acute{K})^T \cdot 1_L\right), \tag{20}$$

Fig. 2 shows the block diagram of the FAVOR+ mechanism.

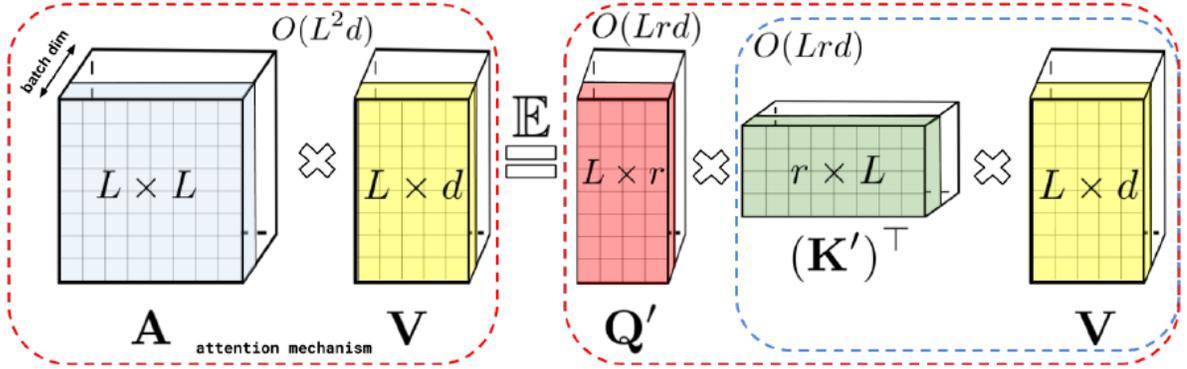

Fig. 2. The block diagram of FAVOR+ mechanism (Choromanski *et al.*, 2020)

**3.4 Bidirectional Long Short-Term Memory (BiLSTM)**

LSTM networks have a significant advantage over traditional RNNs due to their ability to capture long-term dependencies in sequential data. Traditional RNNs suffer from the vanishing gradient problem, which makes it difficult for them to learn and propagate context information across long sequences. This limitation is elegantly addressed by LSTMs, which incorporate a memory cell and gating mechanisms. These features allow LSTMs to control and manage the flow of information, deciding what to retain and what to discard over different time scales. This makes LSTMs particularly effective in tasks were understanding the context and relationships between different parts of the input sequence is crucial, such as in natural language processing, time series analysis, and more (Wang *et al.*, 2023). Therefore, LSTMs often outperform traditional RNNs in tasks involving long sequences and complex dependencies. The operations within an LSTM unit are governed by the equations below:

$$f_t = \sigma(W_f.[h_{t-1}, x_t] + b_f), \tag{21}$$

$$i_t = \sigma(W_i.[h_{t-1}, x_t] + b_i), \tag{22}$$

$$\tilde{C}_t = tanh(W_C \cdot [h_{t-1}, x_t] + b_C), \tag{23}$$

$$C_t = f_t * C_{t-1} + i_t * \tilde{C}_t, \tag{24}$$

$$o_t = \sigma(W_o \cdot [h_{t-1}, x_t] + b_o), \tag{25}$$

$$h_t = o_t * tanh(C_t), \tag{26}$$

Here $f_t$, $i_t$, $o_t$ represent the activations of the forget, input, and output gates, respectively; $C_t$ is the cell state; $h_t$ is the hidden state; $x_t$ is the input at the current time step; $\sigma$ denotes the sigmoid activation function; and $W$ and $b$ are the weights and biases associated with each gate. These equations collectively enable LSTMs to effectively capture temporal dependencies within data sequences. The block diagram of LSTM is shown in Fig. 3.

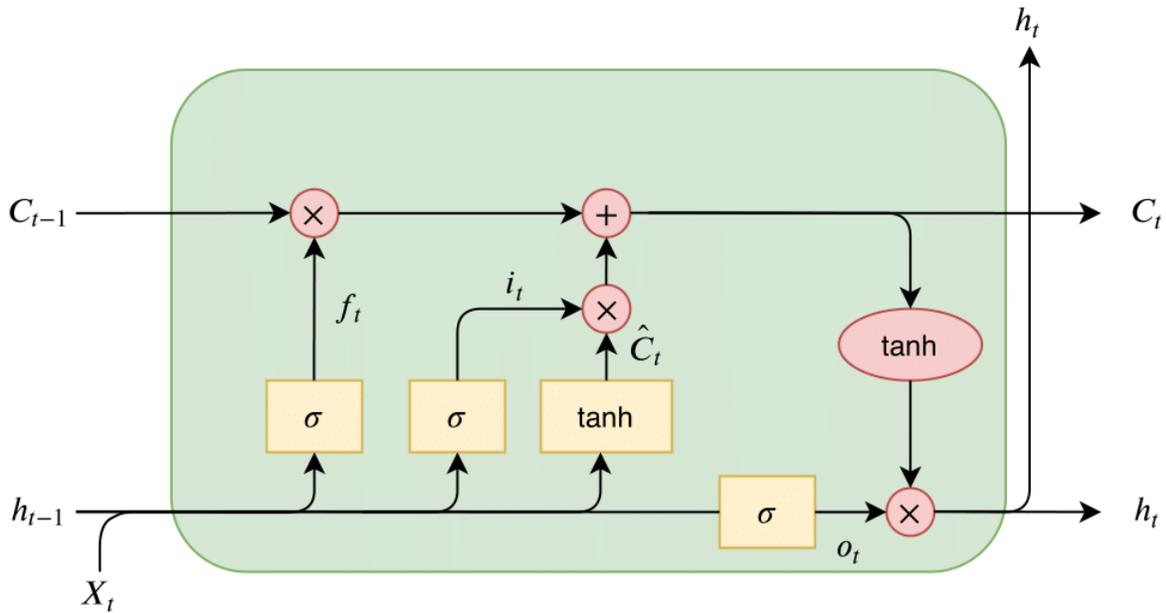

Fig. 3. The block diagram of LSTM

BiLSTM networks are an extension of traditional LSTM networks. BiLSTMs process data in both directions with two separate hidden layers which are then fed forwards to the same output layer.

Unlike a traditional LSTM which processes data sequentially from the beginning to the end of the sequence, a BiLSTM processes data in both directions. One LSTM layer processes the sequence from the start to the end (forward), while other processes it from the end to the start (backward). This bidirectional processing helps capture patterns that may be overlooked by a unidirectional LSTM, as it allows the network to capture information from future states in addition to past states. Mathematically, the output of a BiLSTM at a given time step t is typically represented as the concatenation of the forward hidden state and the backward hidden state. If we denote the forward and backward hidden states at time as $t \rightarrow h_t$ and $\leftarrow h_t$ respectively, the output $y_t$ at time $t$ can be computed as:

$$y_t = [\vec{h}_t, \overleftarrow{h}_t] \tag{27}$$

This output $y_t$ is then passed to the next layer in the network or used to compute the prediction for the current time step. The ability to capture both past (backward) and future (forward) context makes BiLSTM a powerful model for tasks that require understanding the entire context of the input sequence, such as language modeling, text generation, and machine translation (Wang *et al.*, 2023).

**3.5) The Proposed Performer Neural Network**

According to the previous subsections, in this subsection the proposed Performer is introduced. In this structure, after performer blocks, two BiLSTM and FC layers are added to increase the performance of the prediction. The block diagram of LSTM is shown in Fig. 4.

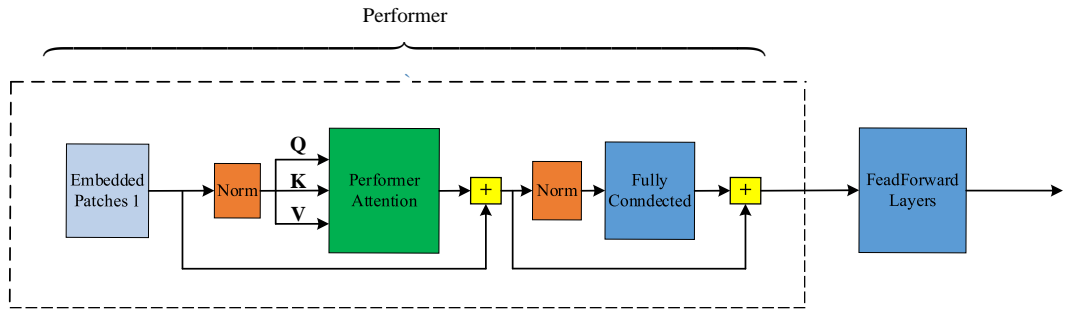

Fig. 4. The block diagram of the Perfomer neural network

In the structure you mentioned, two BiLSTM and FC (Fully Connected) layers are added after the Performer blocks to increase the performance of the prediction. The integration of BiLSTM in the feedforward of the Performer enhances the model's performance by combining the strengths of both models. The Performer can capture long-range dependencies efficiently, while the BiLSTM can process sequential data effectively. This combination allows the model to handle a wider range of data patterns, improving its predictive performance. The absence of a decoder suggests that this model is primarily designed for tasks such as feature extraction, representation learning, or classification rather than sequence-to-sequence tasks that require an explicit decoder.

**4) Simulations**

This section delves into the intricate procedures involved in training and validating the proposed Performer model. This model is specifically designed to predict the price fluctuations of Bitcoin, Ethereum, and Litecoin span from January 2018 to March 2024 across various time intervals, including daily and hourly. The models used for comparison include BiLSTM, Multi-head Attention Transformer with and without technical indicators, Performer, Performer integrated with BiLSTM. These models are evaluated using four performance metrics: Mean Squared Error (MSE), R-square, Root Mean Squared Error (RMSE), and Mean Squared Logarithmic Error (MSLE). These metrics are employed to demonstrate the comparative performance of the proposed methods. Furthermore, the proposed networks are juxtaposed with other methods cited in the current state of the art, providing a comprehensive comparison of their effectiveness.

**4.1) Effectiveness of the Price Prediction Methods in terms of MSE**

In this subsection, $RMSE$ is introduced to deeply evaluate the performance of the neural networks. Equ. (28) shows the RMSE formula.

$$RMSE = \sqrt{\frac{1}{n}\sum_{i=1}^{n}(y_i - \hat{y}_i)^2} \quad , \tag{28}$$

where $y$ is the actual value and $\hat{y}$ is the predicted value, $i$ is the index of data, and $n$ is the amount of data. $RMSE$ represents the square root of the second sample moment of the differences between predicted values and observed values or the quadratic mean of these differences. It measures the average magnitude of the errors without considering their direction.

The models being evaluated for Bitcoin's daily and hourly timeframe LSTM (Awoke *et al.*, 2021), Stochastic Neural Network (Jay *et al.*, 2020), Hybrid LSTM and GRU (Girsang and Stanley, 2023), BiLSTM, Multi-head Transformer both with and without technical indicators, Performer, and Performer combined with BiLSTM. The Figs. 5-6 shows the performance of the model in terms of the RMSE metric.

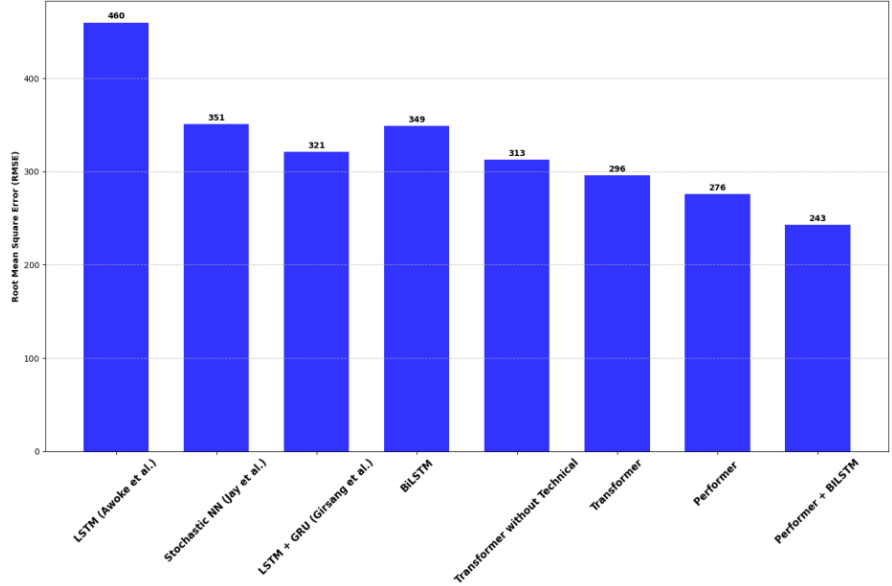

Fig. 5. Comparison of RMSE values for hourly BTCUSD price prediction using state-of-the-art methods and the proposed Transformer-based methods.

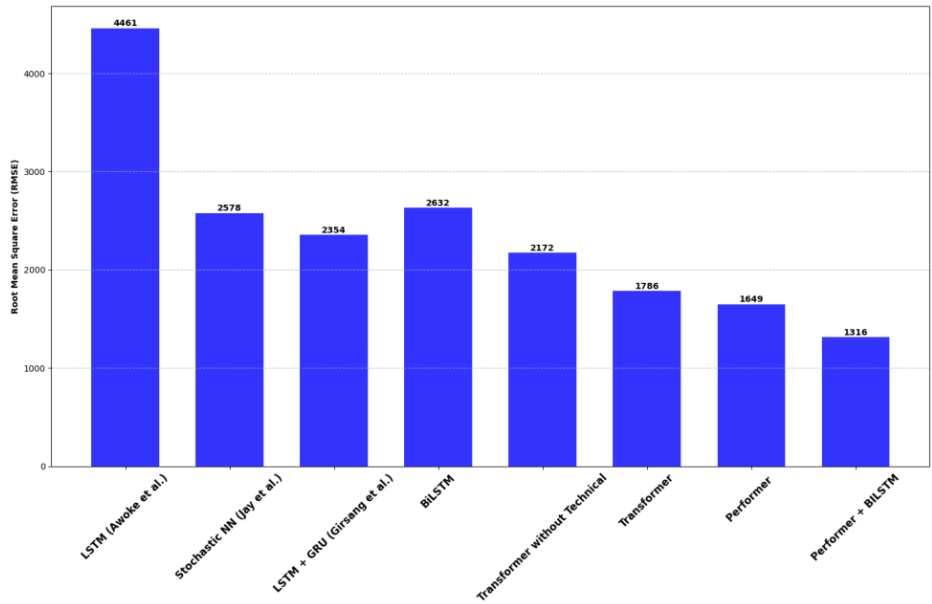

Fig. 6. Comparison of RMSE values for daily BTCUSD price prediction using state-of-the-art methods and the proposed Transformer-based methods.

Based on Figures 5 and 6, it can be inferred that the proposed Transformer-based neural network outperforms other deep learning methods. The Multi-head attention Transformer is better than BiLSTM because it can capture dependencies in the data regardless of their position in the sequence. This is particularly useful in tasks where the position of the data points is not as important as their relationship to each other. On the other hand, BiLSTM, being a recurrent model, processes data sequentially, which can make it less efficient at capturing long-range dependencies. Moreover, technical indicators provide valuable information about market trends and patterns. When used as feature extractors, they can enhance the model's ability to capture and learn from these patterns, leading to improved performance. These indicators can provide insights into various aspects of the market, such as trend direction, volatility, momentum, and market strength, which are crucial for predicting price movements. By incorporating this information into the model, we can improve its ability to make accurate predictions.

In the original Transformer model, the Multi-head attention mechanism computes a weighted sum of all input elements for each output element. This requires a quadratic amount of computation and memory with respect to the sequence length, which makes it difficult to process long sequences. The Performer, on the other hand, uses a technique called Random Feature Maps to approximate the attention mechanism. This technique allows the Performer to compute the attention mechanism in linear time, which significantly reduces the computational cost and allows the model to scale to much larger sequences. Moreover, the Performer maintains a similar level of expressiveness as the original Transformer. It can model complex patterns in the data and capture long-range dependencies between elements in the sequence, which is crucial for many tasks such as language modeling and time series prediction.

The integration of BiLSTM in the feedforward of the Performer enhances the model's performance by combining the strengths of both models. The Performer can capture long-range dependencies efficiently, while the BiLSTM can process sequential data effectively. This combination allows the model to handle a wider range of data patterns, improving its predictive performance. In conclusion, the figure suggests that the Performer + BiLSTM model is the best choice for both daily and hourly Bitcoin price prediction, as it has the lowest RMSE and can capture both the attention and the bidirectional dependencies in the data.

**4.2) Evaluating the Performance of Transformer-Based Models in Cryptocurrency Price Prediction**

In this subsection, we embark on a comprehensive exploration in terms of MSE, $R^2$, RMSE, and MSLE for Bitcoin, Ethereum, and Litecoin close price prediction on daily and hourly timeframe data. Equ. (29) represents MSE formula.

$$MSE = \frac{1}{n}\sum_{i=1}^{n}(y_i - \hat{y}_i)^2 , \qquad (29)$$

MSE is calculated as the sum of the squared differences between the actual and predicted values, divided by the number of data points. Specifically, it is the mean of these squared discrepancies across all observations. MSE serves as a criterion for the optimization of neural networks during the training phase, guiding the adjustment of model parameters to minimize prediction errors. The $R^2$ formula is given by (30).

$$R^2 = 1 - \frac{Var(y - \hat{y})}{Var(y)}, \qquad (30)$$

where $R^2$ Measures the proportion of the variance in the dependent variable that is predictable from the independent variables. It provides an indication of goodness of fit and therefore a measure of how well unseen samples are likely to be predicted by the model. The $MSLE$ is defined as follows.

$$MSLE = \frac{1}{n}\sum_{i=1}^{n}\bigl(log(y_i + 1) - log(\hat{y}_i + 1)\bigr)^2, \qquad (31)$$

*MSLE* Measures the ratio between the true and predicted values. Logarithmic transformation is applied so that errors in predicting large and small values are treated proportionally.

The Figures (7-8) depict the effectiveness of the Transformer Multi-head, Performer, and Performer combined with BiLSTM in predicting the daily and hourly closing prices of Bitcoin. Tables 1-2 present a comprehensive performance evaluation of all the models discussed in the previous subsection.

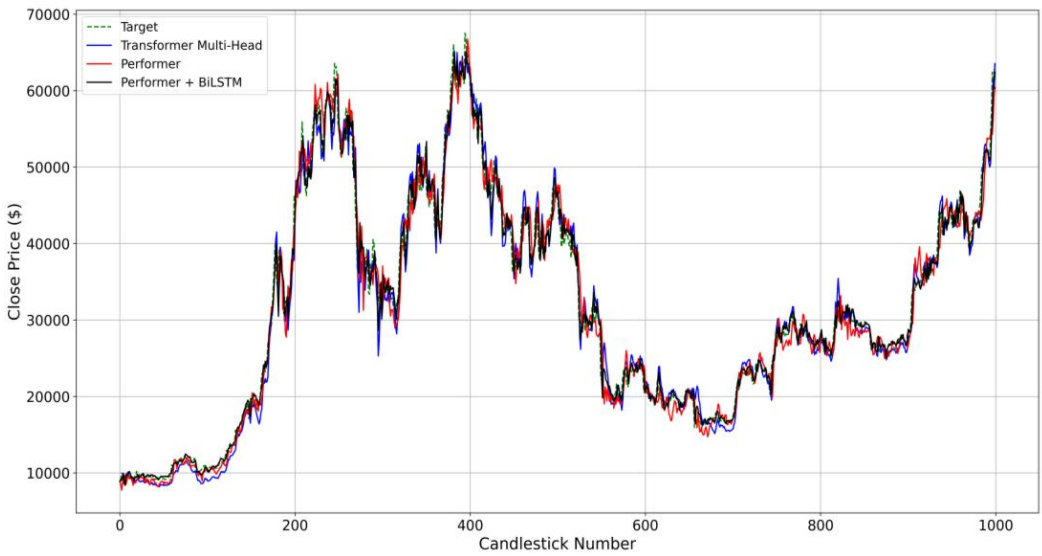

Fig. 7. Price prediction of daily BTCUSD using Transformer Multi-head, Performer, and Performer + BILSTM.

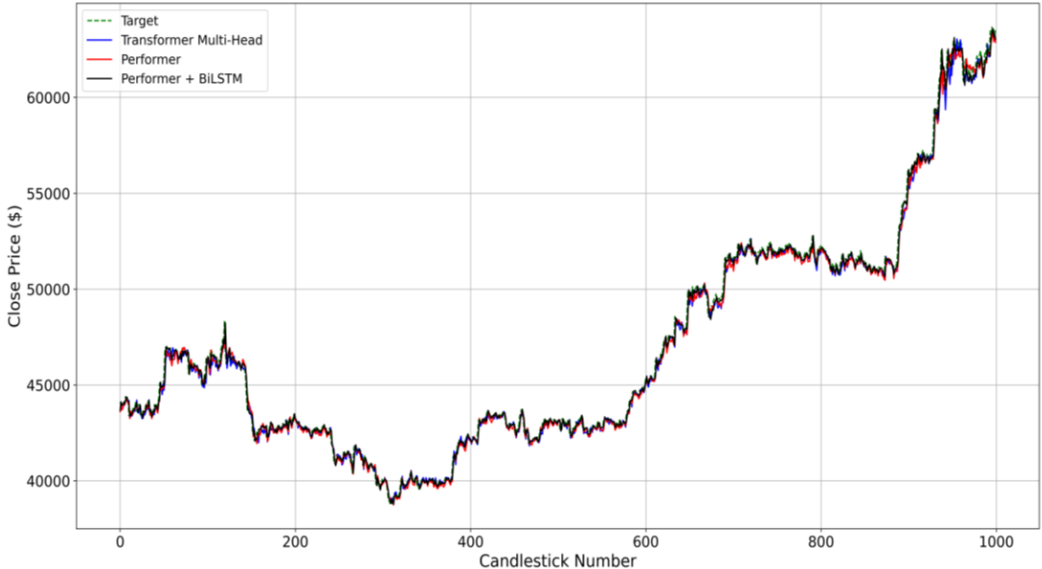

Fig. 8. Price prediction of hourly BTCUSD using Transformer Multi-head, Performer, and Performer + BILSTM.

**Table 1.** Performance indices for price prediction of hourly BTCUSD using some state-of-art methods and proposed Transformer based methods

| Method | MSE | RMSE | R-Square | MSLE |
|---|---|---|---|---|
| LSTM (Awoke *et al.*, 2021) | 195426 | 442 | 0.9971 | 0.00124 |
| Stochastic NN (Jay *et al.*, 2020) | 121802 | 351 | 0.9991 | 0.00076 |
| LSTM+GRU (Girsang and Stanley, 2023) | 103041 | 321 | 0.9992 | 0.00069 |
| BiLSTM | 122312 | 349 | 0.9991 | 0.00075 |
| Transformer Multi-head without Technical Indicator | 98433 | 313 | 0.9994 | 0.00051 |
| Transformer Multi-head | 87710 | 296 | 0.9996 | 0.00032 |
| Performer | 77105 | 276 | 0.9997 | 0.00022 |
| Performer + BILSTM | 59481 | 243 | 0.9998 | 0.00012 |

**Table 2.** Performance indices for price prediction of daily BTCUSD using some state-of-art methods and proposed Transformer based methods

| Method | MSE | RMSE | R-Square | MSLE |
|---|---|---|---|---|
| LSTM (Awoke *et al.*, 2021) | 19900521 | 4461 | 0.897 | 0.0186 |
| Stochastic NN (Jay *et al.*, 2020) | 6646084 | 2578 | 0.918 | 0.0179 |
| LSTM+GRU (Girsang and Stanley, 2023) | 5541316 | 2354 | 0.921 | 0.0172 |
| BiLSTM | 6929056 | 2632 | 0.917 | 0.0179 |
| Transformer Multi-head without Technical Indicator | 4719246 | 2172 | 0.932 | 0.0167 |
| Transformer Multi-head | 3189856 | 1786 | 0.988 | 0.014 |
| Performer | 2720880 | 1649 | 0.99 | 0.008 |
| Performer + BILSTM | 1732590 | 1316 | 0.993 | 0.004 |

The results from Tables (1-2) and Figs. (7-8) show that the Performer + BILSTM model outperforms the other models in predicting the closing price of Bitcoin for both daily and hourly data. The performance of the proposed method for predicting the price of Ethereum are shown in Figs. (9-10) and Tables (3-4).

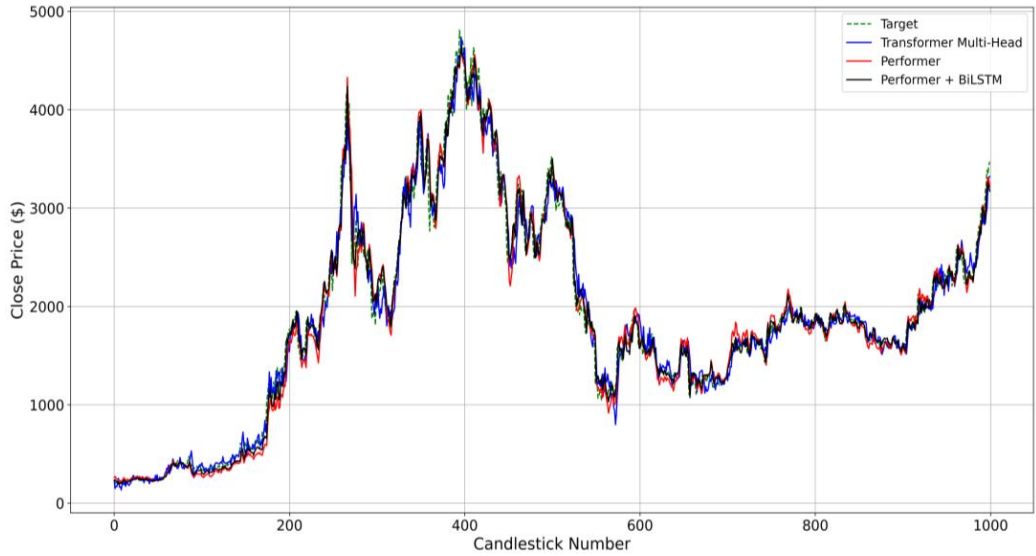

Fig. 9. Price prediction of daily ETHUSD using Transformer Multi-head, Performer, and Performer + BILSTM

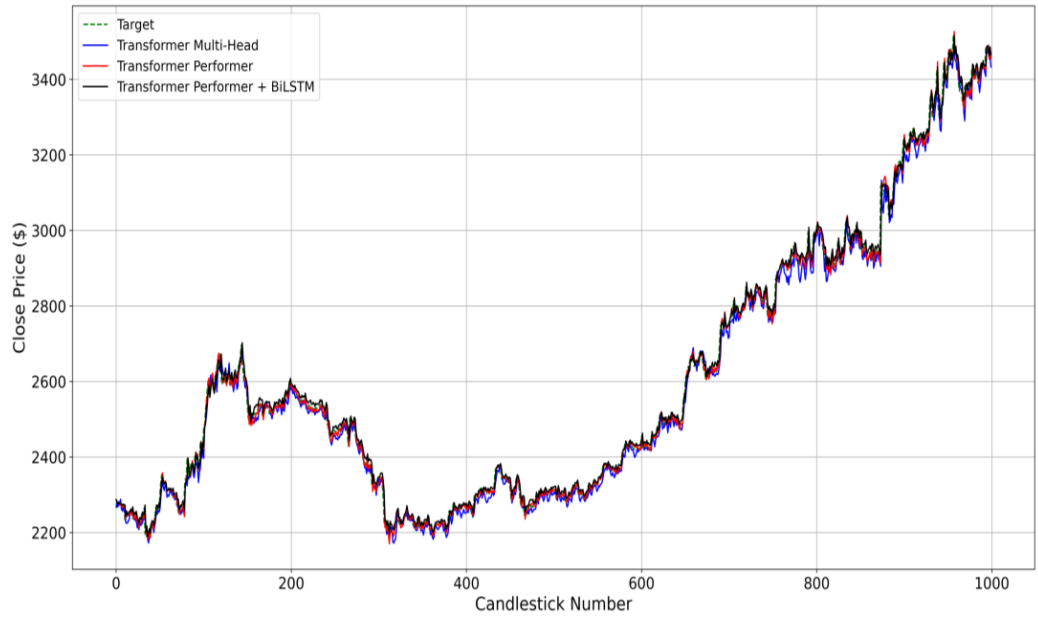

Fig. 10. Price prediction of hourly ETHUSD using Transformer Multi-head, Performer, and Performer + BILSTM

**Table 3.** Performance indices for price prediction of hourly ETHUSD using some state-of-art methods and proposed Transformer based methods

| Method | MSE | RMSE | R-Square | MSLE |
|---|---|---|---|---|
| LSTM (Awoke *et al.*, 2021) | 934 | 30 | 0.9987 | 0.00052 |
| Stochastic NN (Jay *et al.*, 2020) | 798 | 28.2 | 0.9990 | 0.00049 |
| LSTM+GRU (Girsang and Stanley, 2023) | 765 | 27.6 | 0.9991 | 0.00047 |
| BiLSTM | 789 | 28 | 0.9990 | 0.00048 |
| Transformer Multi-head without Technical Indicator | 668 | 26 | 0.9993 | 0.00042 |
| Transformer Multi-head | 518 | 22.7 | 0.9995 | 0.00036 |
| Performer | 477 | 21.8 | 0.9996 | 0.00022 |
| Performer + BILSTM | 386 | 18.3 | 0.9997 | 0.00016 |

**Table 4.** Performance indices for price prediction of daily ETHUSD using some state-of-art methods and proposed Transformer based methods

| Method | MSE | RMSE | R-Square | MSLE |
|---|---|---|---|---|
| LSTM (Awoke *et al.*, 2021) | 38342 | 195 | 0.9698 | 0.048 |
| Stochastic NN (Jay *et al.*, 2020) | 30153 | 173 | 0.9780 | 0.040 |
| LSTM+GRU (Girsang and Stanley, 2023) | 28850 | 170 | 0.9788 | 0.037 |
| BiLSTM | 29601 | 172 | 0.9782 | 0.039 |
| Transformer Multi-head without Technical Indicator | 24807 | 157 | 0.9820 | 0.035 |
| Transformer Multi-head | 14957 | 122 | 0.9890 | 0.023 |
| Performer | 10991 | 105 | 0.9919 | 0.011 |
| Performer + BILSTM | 10017 | 100 | 0.9926 | 0.010 |

The results from Tables (3-4) and Figs. (9-10) show that the Performer + BILSTM model outperforms the other models in predicting the closing price of Ethereum for both daily and hourly data. The performance of the proposed method for predicting the price of Ethereum is shown in Figs. (11-12) and Tables (5-6).

The Performer + BILSTM model achieves better results than the other models in forecasting the closing price of Ethereum for both daily and hourly time frames. The Figs. (11-12) and Tables (5-6) present the performance of the proposed method for predicting the price of Litecoin.

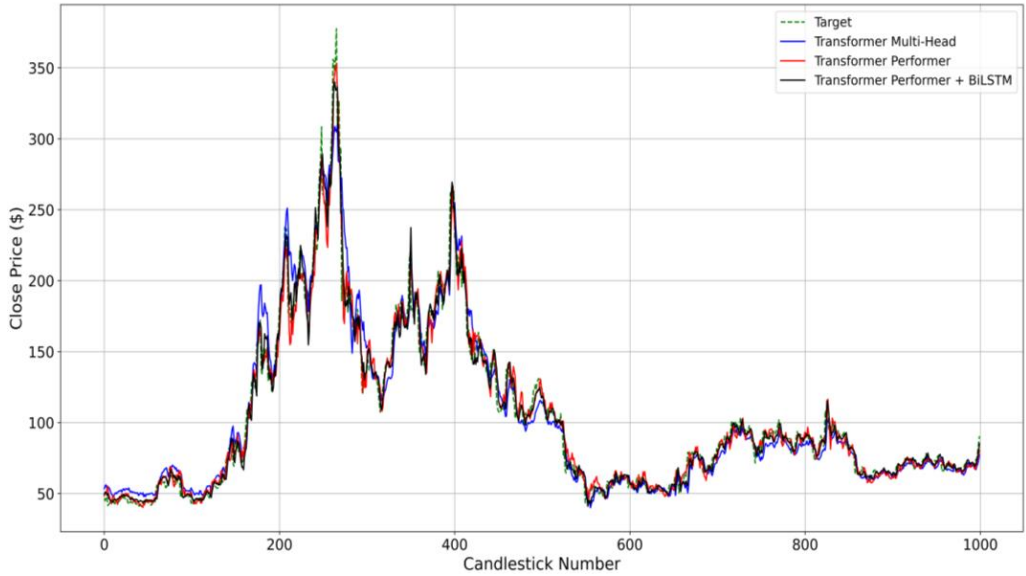

Fig. 11. Price prediction of daily LTCUSD using Transformer Multi-head, Performer, and Performer + BILSTM

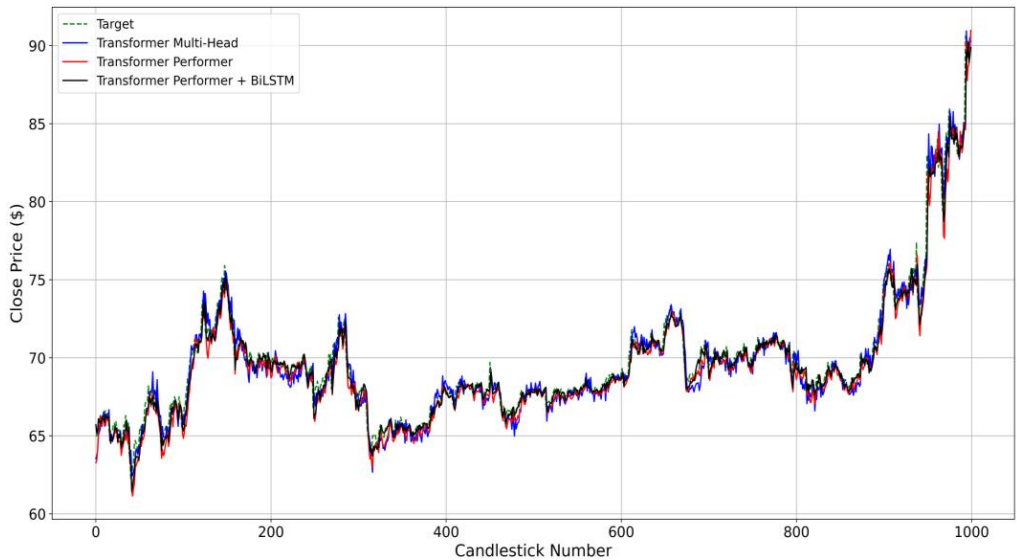

Fig. 12. Price prediction of hourly LTCUSD using Transformer Multi-head, Performer, and Performer + BILSTM

**Table 5.** Performance indices for price prediction of hourly LTCUSD using some state-of-art methods and proposed Transformer based methods

| Method | MSE | RMSE | R-Square | MSLE |
|---|---|---|---|---|
| LSTM (Awoke *et al.*, 2021) | 9.45 | 3.07 | 0.9960 | 0.0067 |
| Stochastic NN (Jay *et al.*, 2020) | 7.74 | 2.78 | 0.9973 | 0.0053 |
| LSTM+GRU (Girsang and Stanley, 2023) | 7.20 | 2.68 | 0.9976 | 0.00053 |
| BiLSTM | 7.85 | 2.80 | 0.9972 | 0.00057 |
| Transformer Multi-head without Technical Indicator | 5.79 | 2.40 | 0.9980 | 0.00045 |
| Transformer Multi-head | 4.96 | 2.22 | 0.9984 | 0.00038 |
| Performer | 4.17 | 2.04 | 0.9987 | 0.00021 |
| Performer + BILSTM | 3.70 | 1.92 | 0.9988 | 0.00020 |

**Table 6.** Performance indices for price prediction of daily LTCUSD using some state-of-art methods and proposed Transformer based methods

| Method | MSE | RMSE | R-Square | MSLE |
|---|---|---|---|---|
| LSTM (Awoke *et al.*, 2021) | 195 | 14 | 0.909 | 0.0192 |
| Stochastic NN (Jay *et al.*, 2020) | 185 | 13.6 | 0.918 | 0.0179 |
| LSTM+GRU (Girsang and Stanley, 2023) | 180 | 13.4 | 0.924 | 0.0177 |
| BiLSTM | 186 | 13.6 | 0.917 | 0.0179 |
| Transformer Multi-head without Technical Indicator | 167 | 12.9 | 0.932 | 0.0167 |
| Transformer Multi-head | 134 | 11.5 | 0.954 | 0.0128 |
| Performer | 65 | 8.1 | 0.980 | 0.0061 |
| Performer + BILSTM | 57 | 7.5 | 0.981 | 0.0057 |

The data from Tables 5-6 and Figures 11-12 demonstrate that the Performer combined with BiLSTM model surpasses other models in forecasting Litecoin's closing price on both daily and hourly timeframes.

## 5) Conclusion

This research has introduced a novel methodology for predicting time series of cryptocurrencies, specifically Bitcoin, Ethereum, and Litecoin. The approach combines the use of technical indicators, a Performer neural network, and BiLSTM to capture temporal dynamics and extract meaningful features from raw cryptocurrency data. The application of technical indicators, such as the RSI and SMA, has allowed for the extraction of complex patterns and trends that might otherwise be overlooked. The Performer neural network, utilizing the FAVOR+, has proven to be more computationally efficient and scalable than the traditional Multi-head attention mechanism used in Transformer models. Furthermore, the incorporation of BiLSTM in the feedforward network has enhanced the model's ability to capture temporal dynamics in the data, processing it in both forward and backward directions. This is particularly beneficial for time series data where past and future data points can influence the current state. The fully connected layers have enabled the model to learn complex non-linear relationships between the features, equipping it to handle the complexity and volatility often seen in cryptocurrency price movements. The proposed method has been applied to the hourly and daily timeframes of major cryptocurrencies such as Bitcoin, Ethereum, and Litecoin, and its performance has been compared with other methods documented in the literature. The results have demonstrated the potential of the proposed method to surpass the predictive performance of existing models, marking a significant advancement in the field of cryptocurrency price prediction.

In future work, we aim to refine the model further and extend its applicability to other financial markets and various types of cryptocurrencies. The ongoing evolution of this research could lead to more precise and efficient predictive models in financial analysis. We also plan to explore the integration of diverse features such as sentiment analysis, social media activity, and news headlines. Additionally, we intend to enhance our method by optimizing the network architecture, fine-tuning the hyperparameters, and applying regularization techniques to mitigate overfitting.